\documentclass{aa}

\usepackage{amssymb}
\usepackage{amsmath}
\usepackage{color}
\usepackage{orcidlink}
\usepackage{natbib}
\bibpunct{(}{)}{;}{a}{}{,}
\usepackage{hyperref}

\bibpunct{(}{)}{;}{a}{}{,}

\usepackage{graphicx}

\usepackage[varg]{txfonts}

\begin{document}

 \title{Maximal efficiency of geometric beaming and polarization of highly super-Eddington X-ray sources}

   \subtitle{}

   \titlerunning{Geometric beaming}

   \author{Varpu Ahlberg,
          \inst{1}\orcidlink{0009-0006-9714-5063}
          Alexandra Veledina,
          \inst{1,2}\orcidlink{0000-0002-5767-7253} 
          Eugene Churazov,
          \inst{3,4} \orcidlink{0000-0002-0322-884X}
          Ildar Khabibullin
          \inst{5,3}\orcidlink{0000-0003-3701-5882}
          }

    \institute{Department of Physics and Astronomy,
              FI-20014 University of Turku, Finland\\
              \email{varpu.a.ahlberg@utu.fi} 
              \and
              Nordita, KTH Royal Institute of Technology and Stockholm University, Hannes Alfv\'ens v\"ag 12, SE-10691 Stockholm, Sweden
              \and
              Max Planck Institute for Astrophysics, Karl-Schwarzschild-Str. 1, D-85741 Garching, Germany
              \and
              Space Research Institute (IKI), Profsoyuznaya 84/32, Moscow 117997, Russia
              \and
              Rudolf Peierls Centre for Theoretical Physics, Department of Physics, University of Oxford, Clarendon Laboratory, Parks Rd, Oxford, OX1 3PU, United Kingdom
             }

   \date{Received XXXX / Accepted XXXX}

  \abstract{
    Geometric beaming is an often invoked explanation for the apparent super-Eddington luminosities of ultraluminous X-ray sources.
    It is a natural consequence of the geometrically thick envelope that is expected to surround the inner accretion disk in the super-Eddington accretion regime; as the radiation can only freely escape through a narrow funnel-shaped cavity, the flux along the funnel axis is enhanced.
    A fraction of the photons will nevertheless escape the funnel cavity at larger angles, reducing the effectiveness of the beaming.
    To explore the limits of geometric beaming, we performed Monte Carlo simulations of conservative radiative transfer in an infinitely optically thick funnel while accounting for polarization.
    We discovered that the beaming becomes inefficient when the funnel opening angle is small $(\lesssim15^\circ$) because the photons tend to remain trapped within the cavity until they reach the upper edge of the funnel where they can escape at larger angles.
    For most of the photons to escape a narrow funnel through a small solid angle, the funnel has to be unrealistically long.
    Consequently, geometric beaming could at most boost the apparent luminosity by a factor of a few dozen in our idealized simulations.
    While larger boost factors on the order of $\sim$100 are possible for funnels with a length of $\gtrsim10^5$ Schwarzschild radii, this requires very narrow opening angles of $\lesssim3^\circ$.
    Additionally, we found that aberration due to mildly relativistic velocities does not significantly change this result.
    The maximal polarization degree of light reflected by the funnel remained on the order of a few percent for face-on sources and $\lesssim10\%$ for edge-on sources. 
   }


   \maketitle

\section{Introduction}

Ultraluminous X-ray sources (ULXs) are extragalactic X-ray binaries (XRBs) that appear to exceed the Eddington luminosity for stellar-mass ($\sim$10$M_\sun$) compact objects \citep[for reviews, see e.g.][]{Kaaret2017, King2023}.
Following the discovery of pulsating ULXs (PULXs) evidently harboring neutron stars, it has become clear that the inferred isotropic luminosity can be greater than the Eddington limit of the compact object by a factor of $\sim$100 \citep{Bachetti2014}.
ULXs offer insight into the poorly understood physics of accretion in the super-Eddington domain, which is also key for understanding the growth of supermassive black hole seeds on cosmic time scales.
The extreme apparent luminosities of ULXs cannot, however, be explained by classical accretion disk models, since the radiation pressure is expected to drive powerful outflows that limit the accretion rate \citep{SS73}.
Greater accretion rates may still be possible due to advection of radiation trapped in the accretion flow, leading to a reduction of the radiative efficiency \citep{Begelman1979, Abramowicz1988}.

A number of effects can enhance the observed luminosity of ULXs beyond the expected Eddington limit \citep[see][for a review]{Middleton2026}. 
Firstly, the true limiting luminosity may be higher.
If ULXs feature magnetars, the extremely strong magnetic field ($\gtrsim$$10^{14}~\mathrm{G}$) can sufficiently reduce the electron scattering opacity of the accreting gas to raise the Eddington limit \citep{Canuto1971, Mushtukov2015}.
This has been used to explain the most luminous PULXs discovered, which exceeded the Eddington luminosity by a factor of $\sim$1000 \citep{Israel2017}.
Nonetheless, the number of accreting magnetars is expected to be small, as none are observed in the Milky Way.
Additionally, if some non-pulsating ULXs feature intermediate-mass black holes, they can produce large luminosities while remaining in the sub-Eddington regime \citep{Colbert1999, Farrell2009}.

Secondly, the radiatively driven outflow or geometrically thick accretion disk of an ULX may allow light to escape only through a funnel-shaped cavity.
This causes the source to be geometrically beamed akin to a searchlight, enhancing the flux along the funnel axis.
Even if the source luminosity is Eddington-limited, the beaming would make the inferred isotropic luminosity appear much greater \citep{King2001}.
In contrast, these sources would appear highly obscured when viewed away from the funnel axis.
Obscured XRBs such as SS433 may thus represent ULXs viewed edge-on \citep{Fabrika2004, Koljonen2020}.
The recently discovered little red dots (LRDs) have also been suspected as highly obscured super-Eddington sources \citep{Narayan2024, King2025, Silk2026}.

In the idealized case for geometric beaming, all of the emitted radiation is assumed to escape through the solid angle subtended by the funnel opening from the central source.
However, this approximation can overestimate beaming efficiency.
General relativistic radiative magnetohydrodynamical (GRRMHD) simulations of super-Eddington accretion show a varying degree of collimation, but a significant fraction of the luminosity escapes away from the funnel opening \citep[see, e.g.,][]{Jiang2014,Kitaki2021,Zhang2025}.
This is in part explained by the finite opacity of the outflow allowing radiation to ``leak'' beyond the funnel cavity.
Moreover, a fraction of the radiative luminosity is lost to the kinetic luminosity of the outflow \citep{Sadowski2015}.

In addition to GRRMHD simulations, the geometric beaming in ULXs has been studied using simplified prescriptions of the radiative transfer.
This approach is more suitable for performing parameter studies of the beaming, although the simplifying assumptions for the interaction of photons with the funnel walls may bias the results.
Beaming by a perfectly absorbing and thermally re-emitting funnel was first studied by \citet{Abolmasov2009} using the diffusive approximation and more recently by \citet{Karmakar2025} using a Monte Carlo (MC) model.
A model for a fully ionized, perfectly reflecting funnel was first considered in \citet{Dauser2017} using a MC model that approximated the reflection with single electron scattering on a thin surface.
Similarly, \citet{Mushtukov2021} and  \citet{Mushtukov2023} employed an MC model that assumed multiple isotropic scatterings within a semi-infinite funnel.
Given the approximations used in prior works, we were motivated to study the geometric beaming with multiple electron scatterings.
Moreover, we wanted to take polarization into account, since X-ray polarimetric observations of ULXs will be possible with the future enhanced X-ray Timing and Polarimetry \citep[eXTP,][]{Zhang2025extp, Zhou2025} mission or the proposed Enhanced X-ray Polarimetry Observatory \citep[EXPO,][]{Soffitta2026} mission.

For this purpose, we developed an MC code that simulates geometric beaming in an infinitely optically thick funnel with either multiple Thomson scatterings or Lambertian re-emission of absorbed photons from the funnel walls.
By using semi-analytical models for the phase functions, we were able to represent the ideal scenario without having to simulate multiple interactions within the funnel.
This made it feasible to calculate the results for a wide parameter range.
Accordingly, we made a comprehensive table of ideal boost factors and derived an approximate analytical expression for the face-on boost factor.
In Sect. \ref{sect:Models}, we outlined the development of our model, in Sect \ref{sect:results}, we performed a parameter study.
Finally, in Sect. \ref{sect:discussion}, we discuss our findings and summarize them in Sect. \ref{sect:conclusions}.

\section{Models} \label{sect:Models}

\subsection{Geometry and polarization}

\begin{figure}
\begin{center}
\includegraphics[width=\linewidth]{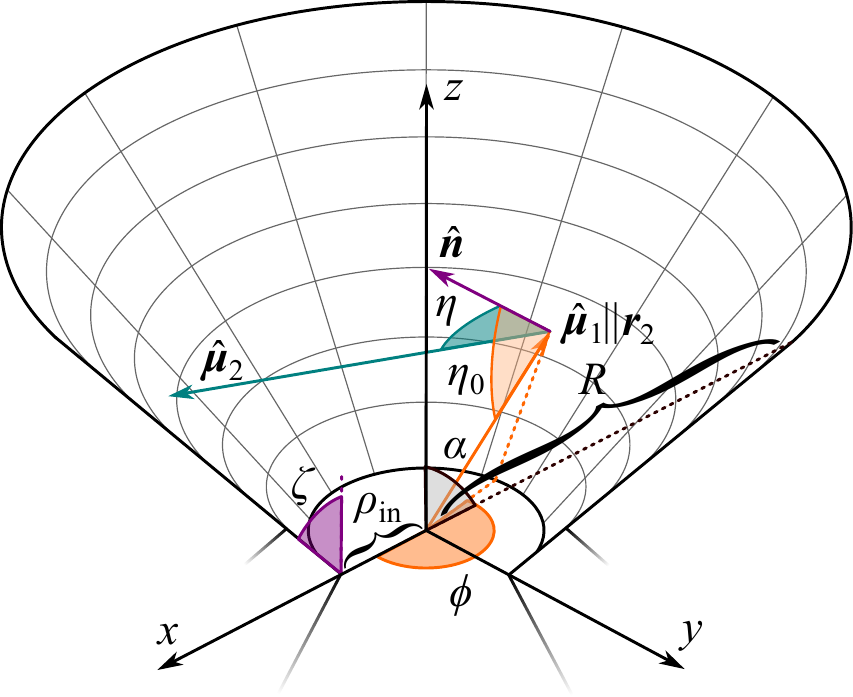}
\end{center}
\caption{Illustration of the funnel geometry. The incident light originates from the center of the funnel; hence, for the first scattering, $\Vec{ r}_1 = (0,0,0)$ and $\Vec{ r}_2$ is parallel to $\Vec{\hat \mu}_1$.}
\label{fig:geometry}
\end{figure}

We represented the funnel walls as the surface of a conical frustum.
Similar to \citet{Veledina2024a}, we parameterized the geometry using the funnel grazing angle, $\alpha$, and the distance from the center of the funnel to the funnel edge, $R$.
These lead to a funnel height and half-opening angle of $h=R\cos\alpha$ and $\tan \zeta = \left( R \sin \alpha - 1\right)/h$, respectively.
We measured $R$ and all other quantities of distance in dimensionless units of the frustum base radius, $\rho_\mathrm{in}$, which is on the order of a few Schwarzschild radii.
The geometry is shown in Fig. \ref{fig:geometry}.

We used a coordinate system where the $z$-axis is aligned with the symmetry axis of the funnel, $\Vec{\hat \omega} = (0,0,1)$, and where the incident source is located at the origin.
Given the axial symmetry, the alignment of the $x$ and $y$ directions is arbitrary.
In spherical coordinates, we measured the colatitude, $\theta$, from the $z$-axis, and the azimuth, $\phi$, from the $x$-axis.
The light is initially located at $\Vec{r}_1 = (r_{1,x}, r_{1,y}, r{_{1,z}})$ and propagates in the direction of $\Vec{\hat \mu}_1 = (\mu_{1,x}, \mu_{1,y}, \mu_{1,z})$.
It is then reflected or re-emitted at $\Vec{r}_2 = (r_{2,x}, r_{2,y}, r{_{2,z}})$ in a new direction of $\Vec{\hat \mu}_2 = (\mu_{2,x}, \mu_{2,y}, \mu_{2,z})$.
We calculated the point $\Vec{r}_2$ by solving the intersection of the light trajectory with the funnel walls (see Appendix \ref{sect:trajectory}).
Then, we replaced the initial location and direction with the new ones, and continued to propagate the light until it exits the funnel.

The normal of the funnel at an azimuth of $\tan \phi = r_{2,y}/r_{2,x}$ is $\Vec{\hat n} = (-\cos\phi\cos \zeta, -\sin \phi \cos \zeta, \mathrm{sgn}(r_{2,z})\sin \zeta)$.
The funnel material flows parallel to the funnel walls in the direction $\Vec{\hat \beta} = (\cos \phi\sin \zeta, \sin \phi \sin \zeta,  \mathrm{sgn}(r_{2,z})\cos \zeta)$ with a velocity of $\beta=v/c$.
Note that the $z$-component depends on the sign of $r_{2,z}$, since the frustum is mirrored with respect to the $x$-$y$ plane.
The photon direction in the rest frame of the funnel walls can be calculated with the relativistic velocity-addition formula,
\begin{align}
    \Vec{\hat \mu}' &= \delta \left[ \Vec{\hat \mu} - \gamma\beta \Vec{\hat \beta} + (\gamma-1)(\Vec{\hat \mu} \cdot \Vec{\hat \beta}) \Vec{\hat \beta}\right], \\
    \delta &= \frac{1}{\gamma\left[1-\beta (\Vec{\hat \mu} \cdot \Vec{\hat \beta})\right]},\\
    \gamma&=\left(1 - \beta^2\right)^{-1/2},
\end{align}
while the inverse transform from the rest frame to the global frame can be performed by changing the sign of $\beta$.
The cosine of the angle that the light impinging on the walls makes with the normal is $\eta_0 = -\Vec{\hat n} \cdot \Vec{\hat \mu}_1$, and the cosine of the angle between the direction of the emerging light and the normal is $\eta = \Vec{\hat n} \cdot \Vec{\hat \mu}_2$.
In the local frame, these are $\eta_0'=\delta\eta_0$ and $\eta' = \delta \eta$.

To calculate $\Vec{\hat \mu}_2'$ from a given $\eta'$, we used a secondary coordinate system where the local $z'$-direction is aligned with the surface normal, $\Vec{\hat a}_z = \Vec{\hat n}$ (see Fig. \ref{fig:basis}).
The local $x'$-axis is opposite to the projection of $-\Vec{\hat \mu_1'}$ on $\Vec{\hat n}$, $\Vec{\hat a}_x = - (-\Vec{\hat \mu}_1' - \eta_0' \Vec{\hat a}_z)/\sqrt{1 - \eta_0'^2}$.
Note that this is constructed so that the local azimuth of the incoming light, $\varphi_0$, is zero in accordance with the coordinate system used by \citet{Chandrasekhar1960}.
The local $y'$-direction is obtained by the cross product of the other two basis vectors: $\Vec{\hat a}_y = \Vec{\hat a}_z \times \Vec{\hat a_x}$.
In this vector basis, the outgoing direction in the local rest frame with a given azimuth $\varphi$ relative to  $\Vec{\hat a}_x$ is $\Vec{\hat\mu}_2' = \sqrt{1-\eta'^2}(\Vec{\hat a}_x \cos \varphi + \Vec{\hat a}_y \sin \varphi) + \Vec{\hat a}_z \eta' $.

We measured the linear polarization of each photon packet using the normalized Stokes vector $(q,u)$.
From it, one can find the polarization degree (PD) as $\sqrt{q^2 + u^2}$.
We defined a global polarization basis by the projection of $\Vec{\hat \mu}$ on $\Vec{\hat \omega}$, and a local basis by the projection of $\Vec{\hat \mu}$ on $\Vec{\hat n}$.
The transformation between the two bases, including the rotation of the basis induced by the Lorentz transformation, is performed by rotating the Stokes vector with the matrix
\begin{equation}
    \mathbf{R}(\chi) = \left(\begin{matrix} \cos 2\chi & \sin 2\chi \\ -\sin 2\chi&\cos 2\chi\end{matrix}\right),
\end{equation}
where $\chi$ is the rotation angle.
We detail all the relevant rotations in Appendix \ref{sect:rot}.
Since our geometry is axisymmetric in the global polarization basis, the total Stokes $u$ of the escaping emission will be zero.
In this instance, the direction of the polarization is determined by the sign of the Stokes $q$: parallel with the funnel axis when $q>0$, and perpendicular to it when $q<0$.

\subsection{Luminosity and incident emission}
 
The luminosity of a standard advection-free thin accretion disk with a super-Eddington mass supply can be approximated by \citep{Poutanen2007}
\begin{equation}
    L = L_\mathrm{Edd} \left[ 1+\frac{3}{5}\ln( \dot{m}) \right],
\end{equation}
where $L_\mathrm{Edd}$ is the Eddington luminosity and $\dot{m}$ is the mass supply rate in Eddington units.
The logarithmic term corresponds to the emission from the outer disk where the local luminosity does not exceed the Eddington limit.

In a geometrically beamed source, the emission can freely escape only through the funnel opening with a grazing angle of $\alpha$.
An observer located at an inclination angle of $i$ who assumes that the source is isotropic would infer an apparent luminosity of 
\begin{equation}
    L_\mathrm{app}(i) = 4\pi D^2 F(i)= \mathcal{B}(i) \int F(i') \, \mathrm{d\Omega},
\end{equation}
where $\mathcal{B}$ is the boost factor, which is the inverse of the beaming factor defined in \citet{King2001}.
The maximum of the geometric boost factor averaged over inclinations of $i\leq\alpha$ is achieved when all photons escape through the solid angle subtended by the funnel opening from the central source:
\begin{equation}
    \mathcal{B}_\mathrm{max} = \left( 1-\cos \alpha\right)^{-1}.
\end{equation}
Even though the angle-averaged boost factor is limited by $\mathcal{B}_\mathrm{max}$, the boost factor may be greater at some inclination.
Nevertheless, if the flux is fairly even over the opening angle, $\mathcal{B}_\mathrm{max}$ is an approximate upper limit for the boost factor regardless of inclination.

We treated the origin of the incident emission as a point source located at the center of the funnel.
The shape of the emitting region is unimportant as long as its extent is small compared to the vertical size of the funnel.
We considered the emission isotropic and unpolarized by default, but allowed for an arbitrarily beamed incident source following a modified version of the Henyey-Greenstein phase function \citep{Henyey1941}:
\begin{equation}
    F_\mathrm{HGm}(\theta, g) = F_0\frac{(1+g)(1-g^2)}{\left( 1+g^2 - 2g \cos 2\theta\right)^{3/2}},
\end{equation}
where $F_0$ is the angle-averaged incident flux between $0\leq\theta\leq\pi/2$ and $-1 > g > 1$ is the asymmetry parameter. 
The mean angle of the emitted light above and below the $x$-$y$ plane are $\left[\cos \theta\right]_\pm=\pm 0.5(g+1)$, respectively.
The emission is accordingly isotropic when $g=0$, fully beamed upward and downward when $g=1$, and fully beamed to the sides when $g=-1$.

As a more physically motivated emission pattern, we used the approximate angular distribution for an electron-scattering-dominated atmosphere to represent an accretion disk \citep{ Chandrasekhar1960, Sobolev1963, Viironen2004}:
\begin{equation}
    F_\mathrm{es}(\theta) = 2 F_0 |\cos \theta| \frac{1+2.06 |\cos \theta|}{1+(2/3)2.06}.
\end{equation}
The polarization of this emission is
\begin{equation}
    q_\mathrm{es}(\theta) = -11.7 \% \left( \frac{1-|\cos \theta|}{1+3.582 |\cos \theta|}\right),
\end{equation}
where the Stokes $u$ is zero.
Additionally, we also tested the incident emission from a Comptonizing slab as approximated by \citet{Nitindala2025}:
\begin{equation}
    F_\mathrm{comp}(\theta) = 1.73 F_0 |\cos \theta| \frac{1+5.3 |\cos \theta| - 0.2\cos^2\theta}{1 +1.3|\cos \theta| + 4.4 \cos^2 \theta},
\end{equation}
where the polarization is
\begin{equation}
    q_\mathrm{comp}(\theta) =  6.4 \% (1-|\cos \theta|) \frac{1+16.3 |\cos \theta| + 6.2\cos^2\theta}{1 +8.2|\cos \theta| - 2.1 \cos^2 \theta},
\end{equation}
and the Stokes $u$ is again zero.

\subsection{Monte Carlo simulations}

We used an MC code to simulate the radiative transfer in the funnel while keeping the simulations energy-independent as a simplification.
To probe the best-case scenario for geometric beaming, we only considered conservative processes so that all the emitted luminosity exits unabsorbed.
Furthermore, we considered the funnel infinitely optically thick so the light can only escape through the funnel cavity.
However, this is computationally expensive, as the photon packets can undergo an enormous number of scatterings or re-emissions whenever they enter the funnel walls.
Thus, instead of explicitly handling multiple scatterings or re-emissions, we used a phase function for diffuse reflection or emission as the angular distribution of the reflected light.
The phase function is proportional to the amount of photons emitted from a surface element into some solid angle,
\begin{equation} \label{eq:phasefunc}
    \mathrm{d}F = I \eta'~\mathrm{d}\Omega,
\end{equation}
where the dependence on $\eta'$ comes from the projection of the surface area in the emitted direction.
We assumed that the light enters the funnel walls and emerges from it at the same point, disregarding the propagation of the light. 
This approximation is valid as long as the funnel is sufficiently dense that the distance traversed by the light entering the walls is small compared to the size of the funnel.

\begin{figure*}
\centering
\includegraphics[width=17cm]{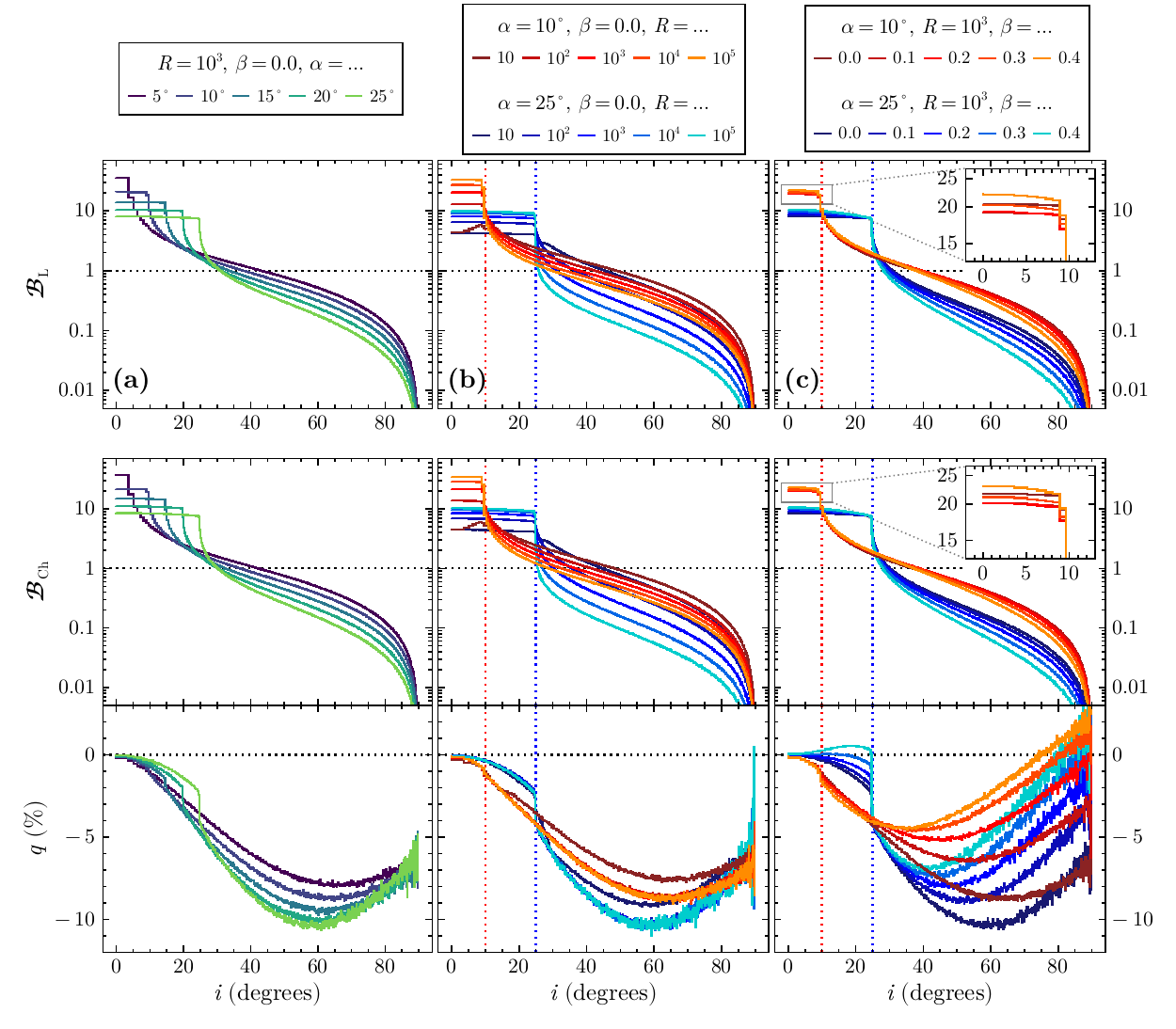}
\caption{The boost factor and normalized Stokes $q$ (PD is $|q|$) as a function of inclination for the Lambertian (top) and Chandrasekhar (bottom) cases. Column (a) shows the dependence of $\alpha$ with fixed $R$, whereas column (b) shows different values of $R$ for two fixed values of $\alpha$, and (c) shows the impact of an increasing outflow velocity $\beta$ for fixed $\alpha$ and $R$.}
\label{fig:parameter_study}
\end{figure*} 

\subsection{Isotropic re-emission from a semi-infinite funnel} \label{sect:Lambert}

First, we considered the case of a fully absorbing funnel that conservatively re-emits all photons at the surface.
We modeled this with the Lambertian phase function, which assumes a constant intensity, and thus depends solely on the angle between the emerging light and the surface normal, $\eta'$ \citep[p.41]{Lambert1760}.
This can represent the angle-dependent bolometric luminosity of a thermally re-emitting surface.
Since the Lambertian phase function is applicable for both emission and reflection, this model also represents an approximate case for diffuse reflection.

The probability density of $\eta'$ for the Lambertian phase function is $P(\eta') = 2\eta'$.
By using the inverse transform method, one can sample values from the distribution with $\eta'=\sqrt{\xi_1}$, where $\xi_1$ is a random number uniformly distributed between 0.0 and 1.0.
As the distribution is independent of the azimuth angle, the outgoing azimuth angle $\varphi$ can be sampled as $2\pi \xi_2$, where $\xi_2$ is a different uniform random number.
The Lambertian model produces zero polarization as it assumes isotropic scattering or emission.

\subsection{Reflection from a semi-infinite electron scattering funnel} \label{Sect:Chandrasekhar}

To represent reflection from a very optically thick and purely scattering funnel, we employed the exact solution for the diffuse reflection from a semi-infinite electron scattering atmosphere first presented in \citet{Chandrasekhar1947}.
This approach correctly handles multiple conservative electron scatterings, including polarization, making it a suitable approximation for reflection from a fully ionized funnel.
Nevertheless, the conservative approximation ignores the recoil effect, which may be significant given the large number of scatterings.
It, however, represents an upper limit for the geometric beaming, as the photons do not lose energy.

The exact solution for the intensity of the reflected light is \citep[Eqs. 164-167]{Chandrasekhar1960} 
\begin{equation}
    \Vec{I}_\mathrm{refl} = \frac{1}{4\eta'} \mathbf{S}(\eta', \varphi, \eta_0', \varphi_0) \Vec{F},
\end{equation}
where $\mathbf{S}$ is the scattering matrix and $\Vec{F}=F(1, q, u)$ is the incident Stokes flux.
The scattering matrix depends on four $H$-functions that do not have closed-form solutions.
To ensure numerical accuracy, we evaluated them using the method described in Appendix \ref{sect:hfunc} within an accuracy of $\lesssim10^{-9}$ and linearly interpolated the results over a grid of 100 equally spaced points.
The polarization basis in Chandrasekhar's solution is defined by the projection of the surface normal; thus, we transformed the polarization into and out of this local basis using the rotations described in Appendix \ref{sect:rot}.

To find the direction of the light after each reflection, we treated $P(\eta', \varphi) \propto\eta' I_\mathrm{refl} $ as the probability density of the angles $\eta'$ and $\varphi$, and calculated their values via rejection sampling.
With given values of $\eta_0'$ and $\Vec{F}$, we drew random values for $\eta'$ and $\varphi$ from a uniform distribution.
We then drew a value $\xi_3$ uniformly distributed between $0$ and $X$, where $X\geq\eta' I_\mathrm{refl}$, and accepted the sampled values if $\xi_3 < \eta' I_\mathrm{refl}(\eta', \varphi, \eta_0', \varphi_0)$.
We found that $X=1.15$ was sufficient for this distribution, but to accelerate the sampling, we reduced it to $X=0.8$ when $\eta_0' < 0.5$.

\section{Results} \label{sect:results}

\subsection{Parameter study}

We first performed a parameter study of the Chandrasekhar and Lambertian models, assuming unpolarized and isotropic incident emission.
The three key parameters of the models are the length of the funnel, $R$, the grazing angle, $\alpha$, and the velocity of the outflow, $\beta$.
We calculated the angle-dependent boost factor over a range of parameter values for both models, while also calculating the polarization in the Chandrasekhar case.
We traced $6\times10^7$ photons for each set of parameters and tallied them into 500 inclination bins of equal solid angle.

The results are shown in Fig. \ref{fig:parameter_study}.
It is immediately clear that the flux distribution is nearly identical between the Lambertian and Chandrasekhar models.
This stems from the fact that both phase functions primarily depend on the projected area of the funnel walls.
Even though the phase functions are different, electron scattering and isotropic scattering are adequately similar for the dependence on $\eta'$ to be dominant.
Ultimately, this indicates that the bolometric boost factor is comparable regardless of whether the emission is reflected or re-emitted from a highly optically thick funnel.

The upward flux boost significantly increases with reduced grazing angles (Fig. \ref{fig:parameter_study}a), albeit less rapidly than implied by the angular dependence of the geometric limit $\mathcal{B}_\mathrm{max}=(1-\cos\alpha)^{-1}$.
Interestingly, the flux escaping toward large inclinations is increased at small grazing angles, making the beaming particularly inefficient for narrow funnels.
This occurs because the light preferentially exits the funnel walls in the direction of the surface normal, which is almost perpendicular to the direction of the funnel opening.
If the solid angle of the funnel opening is minuscule, the photons rarely escape directly.
They are more likely to escape once they are reflected or re-emitted near the top of the funnel, where the solid angle subtended by the opening is much larger, which allows the photons to escape to the sides.

In the Chandrasekhar case with $\beta=0$, the polarization is always perpendicular to the funnel axis with a PD of $\lesssim 10\%$.
The PD reaches its maximum at an inclination of roughly $i \sim 90^\circ - \alpha$, where the average scattering angle is close to $90^\circ$.
The PD is sharply reduced when the source is face-on ($i<\alpha$), because the unpolarized central source becomes visible and the reflection geometry becomes highly symmetric.
When $\alpha$ increases, the PD decreases when the funnel is seen face-on, but conversely increases when $i>\alpha$.

The face-on boost factor increases as a function of $R$ because the number of times the photons are reflected or re-emitted increases, giving them more opportunities to escape directly upward (Fig. \ref{fig:parameter_study}b).
However, the convergence toward $\mathcal{B}_\mathrm{max}$ is slower at small values of $\alpha$.
For $\alpha=25^\circ$, the boost factor is very close to its characteristic value of $\mathcal{B}_\mathrm{max}=10.7$ at $R=10^4$, yet for $\alpha=10^\circ$, the boost factor is only about $40\%$ of $\mathcal{B}_\mathrm{max}=65.8$ even at $R=10^5$.
The polarization is largely independent of $R$ as long as $R\gtrsim10$, since the reflection geometry converges by $R=100$.

The relativistic aberration caused by a moderate outflow velocity does not strongly increase the boost factor (Fig. \ref{fig:parameter_study}c).
In the case of $\alpha=10^\circ$, the face-on boost factor is counterintuitively decreased with velocities of $\beta\lesssim 0.3$.
This is explained by the aberration decreasing the average number of reflections or re-emissions, hence decreasing the number of opportunities the photons have to escape before reaching the top of the funnel.
Even though the probability of escaping directly upwards is increased, it can remain so small in narrow funnels that the boost is ultimately reduced.
For wider grazing angles, such as $\alpha=25^\circ$, the boost increases monotonically with $\beta$ as the photons can more readily escape.
Due to the aberration, the maximum PD is moved to lower inclinations, but the altered scattering angles lead to an overall reduction of the PD.
At large $\beta$, the polarization can even change direction.

\subsection{Impact of the incident emission}

For an isotropic primary source, the boost factor directly translates into the overestimation of the source luminosity inferred from the observed flux.
We also explored different angular distributions for the incident emission.
Since the incident emission loses importance for photons that undergo numerous reflections or re-emissions, the incident emission should be most consequential for small and wide funnels.
To explore this, we tested both a narrower funnel with $\alpha=10^\circ$ and $R=10^3$ and a wider funnel with $\alpha=25^\circ$ and $R=10$.
We focused only on the Chandrasekhar case, as the Lambertian case produced nearly identical results.
We also calculated the boost factor relative to the angle-dependent incident emission, $\mathcal{B}_\mathrm{rel}$.

\begin{figure}
\begin{center}
\includegraphics[width=\columnwidth,trim=0.25cm 0.4cm 0cm 0.42cm,clip=true]{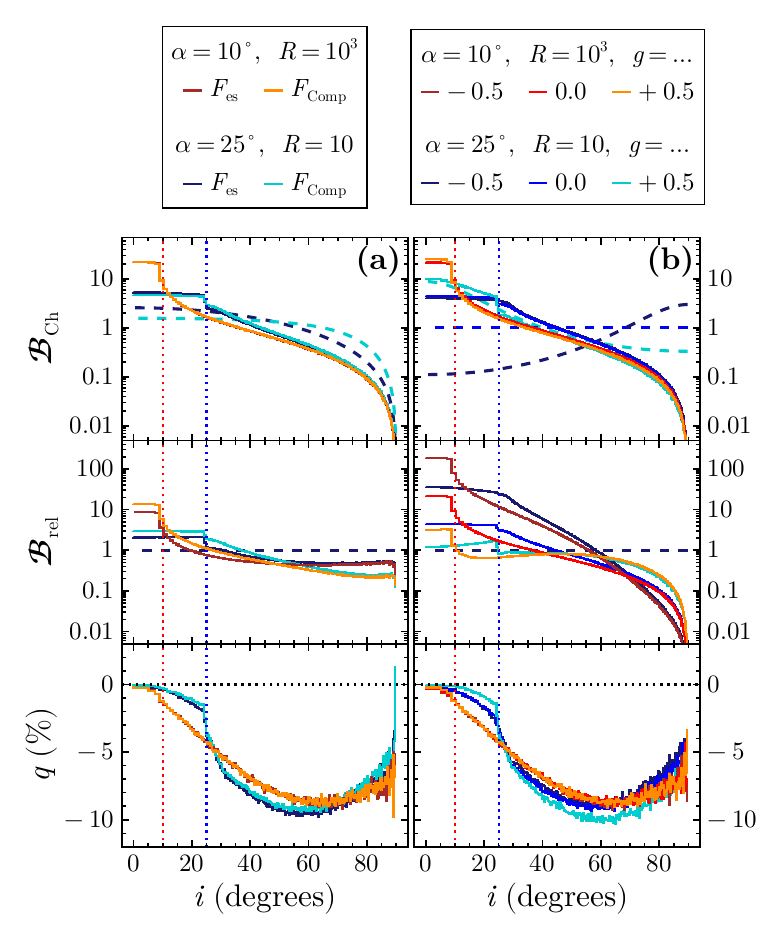}
\end{center}
\caption{Boost factor and Stokes $q$ (PD is $|q|$) as a function of inclination with incident light following either $F_\mathrm{es}(\theta)$ or $F_\mathrm{comp}(\theta)$ (a), or $F_\mathrm{mHG}(\theta)$ for $g=-0.5$ (beamed to the sides), $0.0$ (isotropic), or $+0.5$ (beamed upward) (b).
Both a tall and narrow (reds, $\alpha=10^\circ$, $R=10^3$) and a low and wide (blues, $\alpha=25^\circ$, $R=10$) funnel are shown.
The dotted vertical lines show the opening angles of the funnel, and the dashed lines show the distribution of the incident emission.
The middle panels show the boost factor relative to the incident emission.}
\label{fig:parameter_study_incident}
\end{figure}

First, we compared the incident emissions from an electron-scattering-dominated atmosphere and from a Comptonizing slab, taking into account the polarization.
As shown in Fig. \ref{fig:parameter_study_incident}a, the results are very similar to the isotropic case.
The boost factor at $i\leq\alpha$ is subtly increased from the isotropic case when the funnel is wide, but the narrower funnel shows no clear change.
The polarization of the incident emission also seems to be of little consequence for the final PD, although there is a small difference for the wider funnel.
Since the incident emission is not strongly anisotropic, the boost relative to the incident emission is similar to the boost relative to the total luminosity.

Secondly, we considered an unpolarized source that follows our modified Henyey-Greenstein phase function.
Figure \ref{fig:parameter_study_incident}b shows the results for $g=-0.5$, $0.0$, and $+0.5$.
When beamed to the side (negative $g$), the boost factor at $i\leq\alpha$  is slightly reduced as less light escapes directly upward, but this has no noticeable impact on the polarization.
The relative boost factor, however, is strongly enhanced.
When the source is substantially beamed upward (positive $g$), the face-on boost factor is increased, especially for the wider funnel.
This is entirely caused by an increased amount of light escaping the funnel without interacting with the walls, especially for the wider funnel where $\mathcal{B}_\mathrm{rel}\sim 1$
The increased fraction of escaped direct emission also causes a drop in the polarization at $i\leq\alpha$.
The PD at higher inclinations is somewhat increased in the positive $g$ case for $\alpha=25^\circ$ as the photons are reflected at different angles, but there is no noticeable change for the narrower funnel.

\subsection{Estimating the maximal boost factor}

\begin{figure*}
\centering
\includegraphics[width=17cm,trim=0.25cm 0.4cm 0cm 0.42cm,clip=true]{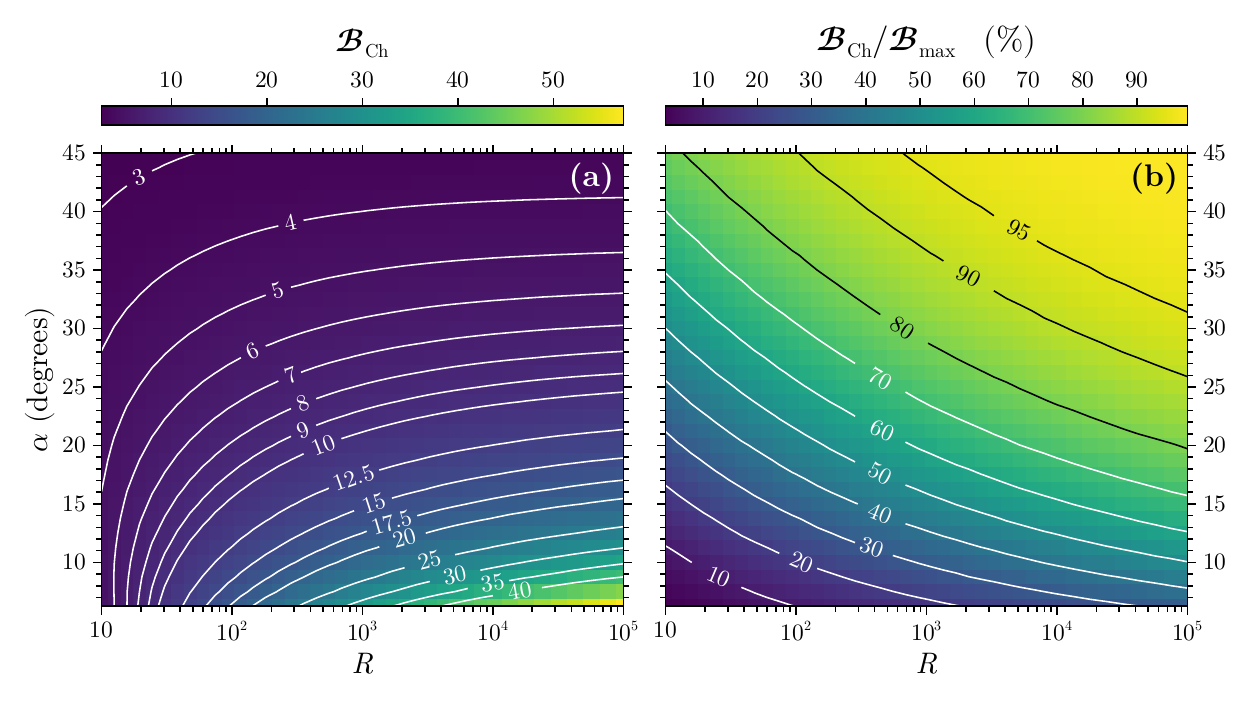}
\caption{Boost factor $\mathcal{B}_\mathrm{Ch}$ for a funnel reflecting according to Chandrasekhar's solution for an electron-scattering atmosphere as a function of $R$ and $\alpha$ averaged over inclinations of $i\leq \alpha$ (a), and the fraction of this $\mathcal{B}_\mathrm{Ch}$ to the geometric maximum of $\mathcal{B}_\mathrm{max}=(1-\cos\alpha)^{-1}$ (b). }
\label{fig:boost_table}
\end{figure*} 

\begin{figure}
\begin{center}
\includegraphics[width=\linewidth,trim=0.25cm 0.5cm 0cm 0.42cm,clip=true]{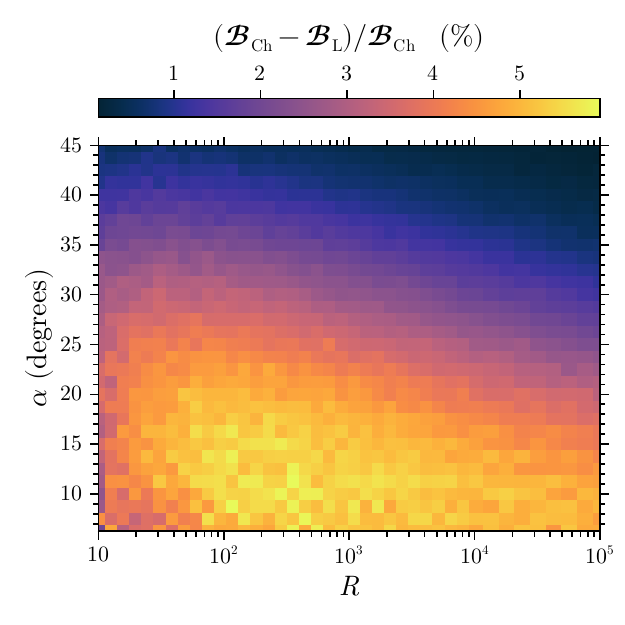}
\end{center}
\caption{The relative difference of the boost factors averaged over inclinations of $i\leq \alpha$ between the Chandrasekhar and the Lambertian model as a function of $R$ and $\alpha$.}
\label{fig:boost_table_comparison}
\end{figure} 

To test the maximal geometric beaming when the source is viewed face-on, we calculated the boost factor averaged over inclinations of $i\leq \alpha$ for both the Lambertian and Chandrasekhar cases.
The boost factor above the funnel opening can be dependent on inclination, but when averaged over the whole opening, it should converge toward $\mathcal{B}_\mathrm{max}$ with increasing $R$.
As can be seen in Fig.~\ref{fig:parameter_study}, the face-on boost factor is nearly inclination-independent as long as $R\gtrsim10$.
We only considered the case of $\beta=0.0$ as the impact of the velocity is secondary to the geometrical parameters for mildly relativistic velocities (see Fig.~\ref{fig:parameter_study}c).

Figure \ref{fig:boost_table}a shows a table of the face-on boost factors for the Chandrasekhar model as a function of $\alpha$ and $R$.
We considered a wide range of grazing angles between $6.25^\circ$--$45^\circ$ and lengths between $10^1$--$10^5$.
Figure \ref{fig:boost_table_comparison} shows the relative difference between the Chandrasekhar and Lambertian models, which remains $\lesssim6\%$ across the whole parameter space.
We did not include lower grazing angles in this table, as the half-opening angle of the funnel becomes negative if $\sin\alpha < 1/R$ ($\alpha <5.7^\circ$ for $R=10$).
Assuming that the funnel base radius is on the order of several Schwarzschild radii and that the scale of the outflow is roughly proportional to $\dot{m}$ \citep{Poutanen2007}, values of $R$ greater than $10^4$ are likely unrealistic, yet we included larger funnel lengths for completeness.

The boost factor increases rapidly with decreasing $\alpha$, but it is much smaller than $\mathcal{B}_\mathrm{max}$ since a significant fraction of the light escapes to the sides.
Furthermore, the boost factor as a function of $R$ converges to this limit extremely slowly at small grazing angles (Fig. \ref{fig:boost_table_comparison}b).
As an example, the boost factor for $\alpha \gtrsim 30^\circ$ reaches $\geq0.9 \mathcal{B}_\mathrm{max}$ by $R=10^4$.
For $\alpha \lesssim 10^\circ$, however, the boost remains less than half of $\mathcal{B}_\mathrm{max}$ even at the extreme funnel length of $R=10^5$.
Empirically, we find that the boost factor can be analytically approximated as
\begin{equation}
    \frac{\ln\mathcal{B}}{\ln\mathcal{B}_\mathrm{max}} \approx  \exp\left(1.14 \frac{\ln \sin \alpha}{\ln R}\right) ,
\end{equation}
which fits the boost factor within our grid of parameters with an accuracy of $\sim$$10\%$.
Although this approximation systematically overestimates $\mathcal{B}$ for small $R$ and underestimates it for large $R$, it demonstrates that the boost becomes less sensitive to the funnel length as $\alpha$ decreases.

A boost factor on the order of $\gtrsim$10 is possible at $\alpha \lesssim10^\circ$ with $R\gtrsim10^2$, yet our parameter range does not include values of $\mathcal{B}\gtrsim100$.
Consequently, we made a few individual simulations with smaller grazing angles to find when greater boost factors are possible.
For both models, we find that $\mathcal{B}\gtrsim100$ is reached at $\alpha \lesssim 2.5^\circ$ when $R=10^4$ and at $\alpha \lesssim 3.5^\circ$ when $R=10^5$.
To reach factors of $\mathcal{B}\gtrsim 200$, one needs $\alpha \lesssim1.0^\circ$ when $R=10^4$ and $\alpha \lesssim1.5^\circ$ when $R=10^5$.

\section{Discussion} \label{sect:discussion}

\subsection{Efficiency of geometric beaming}

Throughout our parameter study, the face-on boost factor in narrow funnels remained much less than the geometric limit of $\mathcal{B}_\mathrm{max} = (1-\cos\alpha)^{-1}$ for any realistic value for the funnel length.
This confirms that assuming a boost factor of $\mathcal{B}_\mathrm{max}$ can significantly overestimate the efficiency of geometric beaming.
Furthermore, the real funnel has a finite optical depth that allows some of the light to leak through the funnel, and the photons can lose some of their energy to the outflow.
Thus, the true boost factor can be even lower than our idealized estimate.

The primary cause for the inefficiency of geometric beaming in our simulations is photons escaping at large angles to the funnel axis.
This is possible even though the funnel is infinitely optically thick, as the solid angle subtended by the funnel opening increases as photons traverse closer to the top of the funnel.
This can particularly impact the geometric beaming when the opening angle is small, since the diminutive solid angle of the opening ($0.004\times4\pi ~\mathrm{sr} $ for $\alpha=5^\circ$) makes the photons unlikely to escape from low in the funnel.
We can estimate the probability that a photon remains inside an infinitely tall funnel after interacting with the walls by integrating the Lambertian phase function over the opening: $P_\mathrm{rem}=1-1/2\pi \int 2\eta ~\mathrm{d}\Omega=1-\sin^3 \alpha$.
Thus, the expected number of reflections or re-emissions prior to escape is $\left(\sum \limits_{N=1}^\infty  N P_\mathrm{rem}^N \right) /\left(\sum \limits_{N=1}^\infty  P_\mathrm{rem}^N \right) = \sin^{-3} \alpha$.

On the other hand, the number of reflections or re-emissions as a function of the funnel height can be approximated using Fick's first law of diffusion in a cylindrical funnel.
Assuming a constant photon flux and diffusivity, the photon number density is $n_\mathrm{p}\propto (h-z) $, where $z$ is the distance along the funnel axis and $h$ is the funnel height. 
The escape probability as a function of $z$ is $P_\mathrm{esc}\propto1-(h-z)/\sqrt{1+(h-z)^2}$, and the average number of reflections or re-emissions before a photon escapes is $\int_0^h n_\mathrm{p} P_\mathrm{esc} \, \mathrm{d}z \propto h^2 - h \sqrt{1+h^2} +\mathrm{arcsinh} \, h \sim \log h$.
While the assumed cylindrical geometry underestimates the dependence on $h$, it demonstrates that the funnel length, $R$, needed for the boost factor to approach $\mathcal{B}_\mathrm{max}$ grows at an immense rate as the opening angle decreases.
This is also evident by the hyperbolic shape of the fractional boost factor contours in Fig. \ref{fig:boost_table}b.

If the funnel absorbs a fraction $f$ of photons during each interaction without re-emitting them, the expected number of interactions prior to absorption is $f^{-1}$.
For most of the photons to exit unabsorbed, the number of reflections or re-emissions should be significantly less than this number.
As a consequence, geometric beaming is only possible when the number of photons is nearly conserved ($f\sim0$).
This particularly affects narrower funnels because the number of interactions prior to escape grows as $\propto\sin^{-3}\alpha$.

\subsection{Factors that may affect the beaming}

We found no significant difference for the boost factor corresponding to the Lambertian and electron scattering models (Fig. \ref{fig:parameter_study}).
The bolometric boost factor should therefore be similar for a perfectly absorbing and re-emitting funnel and a perfectly reflecting funnel.
Our simulations produced very similar results to those shown in \citet{Abolmasov2009, Mushtukov2023} for a similar range of parameters.
However, we found smaller boost factors than \citet{Dauser2017} for funnel opening angles of $\alpha \lesssim 15^\circ$.
A key factor that limits the flux boost is the dependence of the surface phase function on $\eta'$ (Eq. \ref{eq:phasefunc}).
This inhibits photons from emerging from the funnel walls at small angles with the surface.
We tested the code with a fully isotropic and a dipole phase function and found that it increased the boost factor dramatically by up to an order of magnitude.

The details of the angular dependence of the incident source also did not have any major effect on the results (Fig. \ref{fig:parameter_study_incident}).
However, if a large portion of the radiation originates from an extended source, such as a jet within the funnel, the geometry of the emission region may begin to matter.
If one approximates this as a point-like ``lamp'' located high up in the funnel, the length of the funnel at that location would effectively be shorter and its opening angle larger than for the central source.
The boost factor as a function of inclination may therefore appear as a combination of two funnels with different geometric parameters.
This would lead to a lower boost factor than the case where all the incident light is emitted at the funnel base.

We assumed that the funnel cavity is devoid of gas, but an optically thin outflow or jet filling the cavity is seen in many GRRMHD simulations of super-Eddington accretion \citep[e.g.,][]{Sadowski2015, Zhang2025}.
If the outflow velocity is strongly relativistic, it would scatter light preferentially toward the funnel opening.
This would further beam the emission, enhancing the total boost factor when combined with purely geometric beaming by the funnel walls.
Even if the density of the outflow is low, the photons undergo many reflections or re-emissions across the cavity and may thus experience a large total optical depth along the whole trajectory.
Whether this is significant enough depends on the density of the medium, the width of the funnel, and the velocity of the outflow.
Modeling this is beyond the scope of this work.

\subsection{Polarization}

\begin{figure}
\begin{center}
\includegraphics[width=\linewidth,trim=0cm 1cm 0cm 0cm]{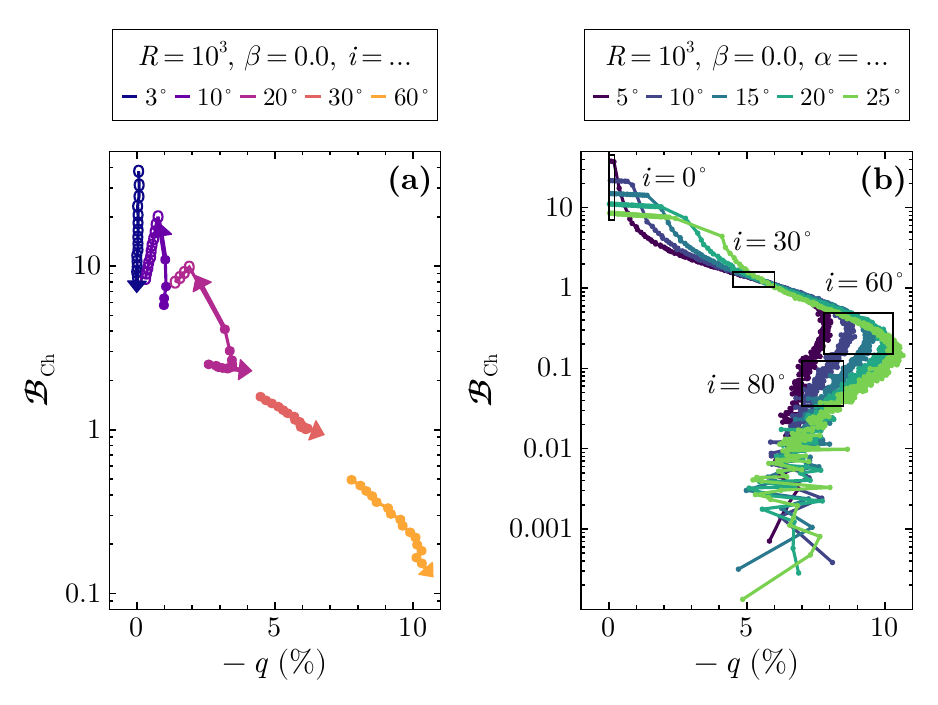}
\end{center}
\caption{The $\mathcal{B}_\mathrm{Ch}$$-$$q$ relation with $\beta=0$ for varying $\alpha$ of 15 equally spaced values between $5^\circ$ and $25^\circ$ with a fixed $i$ (a) and for varying inclination with a fixed $\alpha$  (b). For the former case, the direction of increasing $\alpha$ is marked by arrows, and points for which $i<\alpha$ are marked by open circles. For the latter, regions in the $\mathcal{B}_\mathrm{Ch}$$-$$q$ diagram are labeled with their corresponding inclinations. The funnel length is $R=10^3$. }
\label{fig:b_q_correlation}
\end{figure} 

The PD of the emission reflected from a fully ionized funnel is at most $10 \%$, which is significantly lower than the $\sim$$80\%$ polarization produced by a strongly absorbing funnel where most of the reflected photons are scattered only once \citep{Ahlberg2026}.
However, the unpolarized re-emitted radiation can dilute this strong polarization.
Since the albedo of a medium tends to increase with energy, the Lambertian and Chandrasekhar models may represent the low- and high-energy limits, respectively.
Consequently, the PD should increase with energy.
Nevertheless, if the single-scattered emission dominates over the thermally re-emitted emission at a particular energy band, the PD would be large and energy-independent throughout that range \citep[see][]{Ahlberg2026}.
A full model including the energy-dependent flux and polarization would require taking into account the energy-dependent cross sections and the temperature and density gradients across the whole funnel, which is a complex task.
We leave this to future work.

It may be possible to constrain the observer inclination, the boost factor, or the funnel opening angle from the polarization. 
This can be analyzed using the $\mathcal{B}$--PD relation, which is shown in Fig \ref{fig:b_q_correlation} for the same model setup as Fig. \ref{fig:parameter_study}a. 
At $i< 60^\circ$, both the PD and $\mathcal{B}$ depend strongly on the model parameters, but the polarization becomes insensitive to both $i$ and $\alpha$ at very large inclinations.
While the PD at $i < \alpha$ is only a few percent and hence difficult to detect with high accuracy, a low PD can indicate that a source is beamed toward the observer.
Moreover, the $\mathcal{B}$--PD relation can be used to analyze variability even for face-on sources.
If the opening angle is variable (Fig. \ref{fig:b_q_correlation}a), the PD and the boost factor are correlated as long as $i<\alpha$. 
However, they suddenly become anticorrelated when the opening angle exceeds $i>\alpha$.
This abrupt transition in our model is caused by the sharp edge of the funnel and would be smoother in reality.
Interestingly, there is another sharp turn in the $\mathcal{B}$--PD diagram at $i\gtrsim\alpha+5^\circ$ where the boost factor begins to increase with decreasing $\alpha$ (see Fig. \ref{fig:parameter_study}a).

In contrast, if the opening angle is fixed but the inclination varies due to precession (Fig. \ref{fig:b_q_correlation}b), the PD in the Chandrasekhar model is anticorrelated with the boost factor at low to intermediate inclinations.
Between intermediate inclinations of $30^\circ \lesssim i \lesssim 60^\circ$, the anticorrelation between $\log \mathcal{B}$ and the PD appears nearly linear.
The correlation changes sign at $i \gtrsim 60^\circ$ as the PD begins to steadily decrease while the flux plummets toward zero, although this is also affected by the sharp boundary of our model geometry.
As the dependence of the PD on $i$ and $\alpha$ is opposite for face-on sources, the $\mathcal{B}$--PD correlation may reveal whether a source is precessing or has a variable geometry.
Nevertheless, the opening angle is likely a function of the accretion rate; thus, the intrinsic luminosity can also vary alongside $\mathcal{B}$.

\subsection{The ULX and LRD populations}

The face-on boost factor depends mostly on the funnel geometry, which in turn depends on the accretion rate.
Consequently, if the boost factor of an ULX is possible to determine from the observed luminosity, the accretion rate can be constrained and used in population studies.
By assuming that the funnel is cylindrical and that $\mathcal{B}=\mathcal{B}_\mathrm{max}$, \citet{King2009} inferred the empirical relation $\mathcal{B} \sim \dot{m}^2/73$.
Because geometrical beaming is less efficient for narrower opening angles, this relation can underestimate the accretion rate of more luminous sources.
We find that within the parameter range shown in Fig. \ref{fig:boost_table}, funnels that are large enough that $\mathcal{B} \gtrsim0.9\mathcal{B_\mathrm{max}}$ can at most reach a boost factor of $\sim$10.
Roughly speaking, this indicates that the empirical relation becomes increasingly less reliable when $\dot{m} \gtrsim 30$ if the funnel height is $\lesssim$$10^6$ Schwarzschild radii.

Presuming that LRDs represent massive high-redshift extreme counterparts of edge-on ULXs, or possibly Ultraluminous Supersoft X-ray sources \citep[ULS, ][]{2016MNRAS.456.1859U}, the funnel can be modeled similarly.
The geometrically beamed component consists mainly of hard emission because only the hot inner disk is covered by the funnel.
The distribution of observed X-ray luminosities of LRDs should therefore be dependent on the distribution of the orientation angles and the angular dependence of the flux.
As shown in Fig. \ref{fig:parameter_study}, the visible flux from an obscured source decreases rapidly between about $i\sim\alpha$ and about $i\sim \alpha+10^\circ$, and thus X-ray bright sources should be seldom encountered if the hard X-ray emission is weak at larger inclinations.
Indeed, LRDs are extremely X-ray faint and are generally undetected even in stacked images \citep{Yue2024}, although X-ray emission has been detected from a few individual sources \citep{Fu2026, Hviding2026}. 
Face-on LRDs may have very different spectra than those seen edge-on, yet they would be rare if the funnels are narrow; assuming a random orientation, only $1-\cos \alpha$ of all observers would see the beamed emission.
The funnel opening angle in recent GRRMHD simulations tends to range between $\sim$$15^\circ$ and $\sim$$30^\circ$ at various accretion rates \citep{Kitaki2021, Zhang2025, Fragile2025}, and the X-ray polarization of the funnel-obscured X-ray binary \mbox{Cygnus X-3} implies a narrow opening angle of $10^\circ$ to $15^\circ$.
If these values are representative, about 1--10\% of beamed sources would be viewed face-on.

Our conclusion regarding the maximum beaming factor implies a decreasing contribution of extremely beamed sources to the total radiative output of the source population. 
Given that only the brightest sources, i.e., the most favorably beamed ones, are readily detectable in flux-limited X-ray surveys even with the future NewAthena \citep{2025NatAs...9...36C} and Lynx observatories \citep{2019lxro.rept.....V,2026arXiv260900033L}, one has to be careful in translating the luminosity of resolved sources into the total accretion-powered luminosity of the population.
Analogously to the Soltan argument applied to the cosmic X-ray background \citep{1982MNRAS.200..115S}, the total luminosity is closely connected to the early growth history of supermassive black holes and their initial seed masses.
Another aspect is the impact of radiation from such sources on early pre-heating of the intergalactic medium \citep{2013ApJ...776L..31F,2017AstL...43..211S} and the correlation of X-ray sources with the corresponding 21 cm signal \citep[e.g.,][]{2019MNRAS.489.1127S}.
Our result once again implies that the contribution of extremely beamed sources is unlikely to dramatically change predictions assuming radiation isotropy.
The phenomenological formulae obtained here can be directly used in idealized set-up calculations including photoionization \citep[e.g.][]{2018MNRAS.476.2530S} and as recipes in cosmological simulations with realistic conditions \citep[e.g.][]{2026Natur.657..621C} to verify this in the future.

\section{Conclusions} \label{sect:conclusions}

We investigated the upper limit of the boost factor for geometric beaming in ULXs under idealized conditions of conservative radiative transfer in a semi-infinite funnel.
We performed MC simulations of the problem over an extensive range of parameters, testing both Lambertian re-emission and electron scattering.
We found the beaming to be particularly inefficient in narrow funnels when compared to the geometric limit of $\mathcal{B}_\mathrm{max} = (1-\cos\alpha)^{-1}$, confirming that $\mathcal{B}_\mathrm{max}$ can strongly overestimate the boost factor even in an idealized scenario. 
Our model shows that the highest attainable boost factor with geometric beaming is on the order of ten unless the funnel is very narrow ($\lesssim5^\circ$) or unrealistically large ($\gg 10^4$ Schwarzschild radii).
This suggests that geometric beaming can produce the apparent luminosity of many ULXs, but sources with apparent luminosities of $\gtrsim100 L_\mathrm{Edd}$ should involve some other mechanism to explain the results.
For instance, further collimation may be provided by a relativistic outflow or jet within the funnel cavity.

We found no significant difference between the electron scattering and Lambertian models for the angular dependence of the light emerging from the funnel.
Similarly, modeling the incident source as an electron scattering-dominated atmosphere or a Comptonizing slab also had no noticeable impact compared to an isotropic and unpolarized source.
On the other hand, an arbitrarily beamed incident source could increase the boost factor, purely by allowing more of the light to escape without interacting with the funnel.
For narrow funnels, however, the effect is minor.
Relativistic aberration from an outflowing funnel can potentially enhance the boost factor, but we found that mildly relativistic velocities could counterintuitively reduce the boost factor in narrow funnels.
A sufficiently large velocity can, however, reduce the polarization, or even change the sign of the polarization at certain inclinations depending on the funnel opening angle.

The reflected light from an electron-scattering funnel (in the absence of absorption) had a maximum PD of roughly $5\%$--$10\%$ at grazing angles of $5^\circ$--$25^\circ$.
When viewed face-on, the PD was at most a few percent for narrow funnels, with the polarization increasing as a function of the inclination.
We found that the PD is anticorrelated with the boost factor in a precessing ULX and correlated with the boost factor in a ULX with a variable opening angle.
Future X-ray polarimetry missions, such as the upcoming eXTP or the proposed EXPO, will be sensitive enough to measure these trends.

\begin{acknowledgements}
VA and AV were supported by the Research Council of Finland grants 355672, 372881 and the Centre of Excellence in Neutron-Star Physics (grant 374064).
VA acknowledges support from the Turku University Foundation.
Nordita is supported in part by NordForsk. 
IK was supported by the Simons Foundation via the Simons Investigator Award to A. A. Schekochihin.
\end{acknowledgements}

\bibliographystyle{mnras}
\bibliography{references}

\appendix

\section{The intersect of a conical frustum} \label{sect:trajectory}

The funnel walls lie at a cylindrical radius of $\rho_\mathrm{wall}(z) = 1+|z|\tan \zeta$.
The light starting from point $\Vec{r}_1$ will reach the funnel surface after traveling some distance $d$ along the propagation direction $\Vec{\hat \mu}_1$:  $\Vec{r}_2 = \Vec{r}_1 + d\Vec{\hat \mu}_1$.
The distance $d$ can be found by setting ${r_{2,x}}^2 + {r_{2,y}}^2 = \rho_\mathrm{wall}(r_{2,z})^2$ and solving the resulting equation.
This equation has no positive real solutions when $|\mu_{1,z}|>\cos\zeta$ because the light escapes the funnel without intersecting it.

In two special cases, the problem reduces to a linear equation.
Firstly, when the incident emission is from the origin $\Vec{r_1} = (0,0,0)$, the solution is
\begin{equation}
    d_\mathrm{incident} = \frac{1}{\sqrt{1-\mu_{1,z}^2} -|\mu_{1,z}| \tan \zeta}.
\end{equation}
Secondly, if the trajectory begins on the funnel surface at ${r_{1,x}^2 + r_{1,y}^2} = \rho_\mathrm{wall}(r_{1,z})^2$ and does not cross the $x$-$y$ plane ($r_z$ does not change sign), the solution is 
\begin{equation}
    d_\mathrm{wall} = 2\frac{\mathrm{sgn}(r_{1,z}) \mu_{1,z} \tan\zeta + r_{1,z}\mu_{1,z} \tan^2 \zeta - r_{1,x} \mu_{1,x} - r_{1,y}\mu_{1,y}}{\mu_{1,x}^2 + \mu_{1,y}^2 - \mu_{1,z}^2 \tan^2 \zeta}.
\end{equation}
This solution is always applicable when $r_{1,z}$ and $\mu_{1,z}$ have the same sign.
Otherwise, the light trajectory intersects the funnel surface before crossing the $x$-$y$ plane if $(r_{1,x} -r_{1,z}\mu_{1,x}/\mu_{1,z})^2 + (r_{1,y} - r_{1,z}\mu_{1,y}/\mu_{1,z})^2 >1 $.

Trajectories starting at the funnel surface that cross the $x$-$y$ plane, i.e., for which $r_z/\mu_z <0$ and $(r_x -r_z\mu_x/\mu_z)^2 + (r_y - r_z\mu_y/\mu_z)^2 <1 $, the problem becomes a quadratic equation with only one positive solution as long as $\eta>0$ and $\eta_0>0$:
\begin{align}
    d_\mathrm{wall, c} &= 2\frac{-A + \sqrt{A^2 - 4|r_{1,z}|\tan\zeta(\mu_{1,x}^2 + \mu_{1,y}^2 - \mu_{1,z}^2 \tan^2 \zeta)}}{\mu_{1,x}^2 + \mu_{1,y}^2 - \mu_{1,z}^2 \tan^2 \zeta}, \\
    A &= r_{1,x}\mu_{1,x} + r_{1,y}\mu_{1,y} + \mathrm{sgn}(r_{1,z}) \mu_{1,z} \tan \zeta - r_{1,z}\mu_{1,z} \tan^2 \zeta.
\end{align}
A second positive solution exists if the trajectory goes through the wall, but this does not occur in our model.

\section{Rotating the polarization basis}\label{sect:rot}

\begin{figure}
\begin{center}
\includegraphics[width=\linewidth]{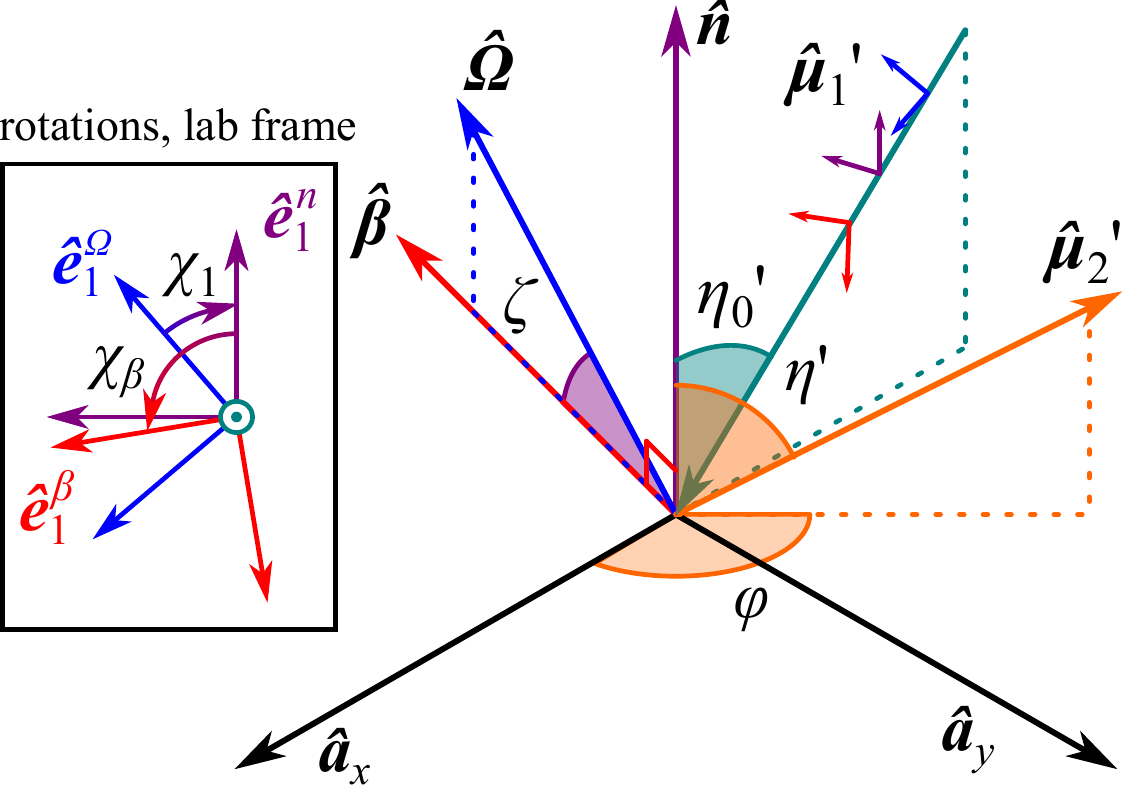}
\end{center}
\caption{Illustration of the geometry and the polarization basis rotation. Note that the rotations are only shown for the lab frame $\Vec{\hat \mu}_1$  }
\label{fig:basis}
\end{figure} 

To calculate the polarization bases and their rotations, we employed the polarization basis formalism of \citet{Loktev2022}:
\begin{equation}
    \mathbf{B}(\Vec{\hat \mu}, \Vec{\hat{l}}) = \left\{\Vec{\hat e}_1^l, \Vec{\hat e}_2^l \right\} =\left\{\frac{\Vec{\hat l} - (\Vec{\hat \mu} \cdot \Vec{\hat l}) \Vec{\hat \mu}}{|\Vec{\hat \mu} \times\Vec{\hat l}|}, \frac{\Vec{\hat \mu} \times\Vec{\hat l}}{|\Vec{\hat \mu} \times\Vec{\hat l}|}, \right\},
\end{equation}
where $\Vec{\hat l}$ is the reference direction for the basis.
The rotation angle between two bases is
\begin{align}
    \cos \chi(\Vec{\hat l}_2 \rightarrow \Vec{\hat l}_1) &= \Vec{\hat e}_2^{l_1} \cdot \Vec{\hat e}_2^{l_2} = \Vec{\hat e}_1^{l_1} \cdot \Vec{\hat e}_1^{l_2},\\
    \sin \chi(\Vec{\hat l}_2 \rightarrow \Vec{\hat l}_1) &= \Vec{\hat e}_2^{l_1} \cdot \Vec{\hat e}_1^{l_2} = -\Vec{\hat e}_1^{l_1} \cdot \Vec{\hat e}_2^{l_2}.
\end{align}
These angles can also be found from the spherical triangle formed by $\Vec{\hat l}_1$, $\Vec{\hat l}_2$, and $\Vec{\hat \mu}$.
Figure \ref{fig:basis} shows an overview of the rotations in our model.

The first rotation in our problem is from the global basis of the incident light, $\mathbf{B}(\Vec{\hat \mu}_1, \Vec{\hat{\omega}})$, to the local one, $\mathbf{B}(\Vec{\hat \mu}_1, \Vec{\hat{n}})$.
This angle is
\begin{align}
    \cos \chi_1 &= \frac{n_\mathrm{z} + \eta_0 \mu_{1,z}}{\sqrt{1 - \mu_{1,z}^2} \sqrt{1-\eta_0^2}}, \\
    \sin \chi_1 &= \frac{\cos \zeta \sin(\psi_1-\phi)}{\sqrt{1-\eta_0^2}},
\end{align}
where $\tan \psi_1 = \mu_{1,y}/\mu_{1,x}$ is the azimuth of the incident direction and $\tan \phi = r_{2,y}/r_{2,x}$ is the azimuth of the surface point.

Then, we calculated the rotation of the basis between the laboratory frame and the funnel rest frame due to relativistic aberration.
The PD is Lorentz invariant, but the PA changes under the Lorentz transform of the electric field vector.
It is convenient to first transform from $\mathbf{B}(\Vec{\hat \mu}_1, \Vec{\hat{n}})$ to $\mathbf{B}(\Vec{\hat \mu}_1, \Vec{\hat \beta})$, since the polarization does not rotate in the Lorentz transform to the basis $\mathbf{B}(\Vec{\hat \mu}_1', \Vec{\hat \beta})$:
\begin{align}
    \cos \chi_\beta &= \frac{-(\Vec{\hat \mu}_1 \cdot \Vec{\hat{n}}) (\Vec{\hat \mu}_1 \cdot \Vec{\hat{\beta}})}{\sqrt{1-\eta_0^2} \sqrt{1-(\Vec{\hat \mu}_1 \cdot \Vec{\hat{\beta}})^2}},\\
    \sin \chi_\beta &= \frac{\mathrm{sgn}(r_{2,z}) (\mu_{1,x} \sin \phi - \mu_{1,y}\cos \phi) }{\sqrt{1-\eta_0^2} \sqrt{1-(\Vec{\hat \mu}_1 \cdot \Vec{\hat{\beta}})^2}}.
\end{align}
Then, a second transform is needed from $\mathbf{B}(\Vec{\hat \mu}_1', \Vec{\hat \beta})$ to $\mathbf{B}(\Vec{\hat \mu}_1', \Vec{\hat n})$:
\begin{align}
    \cos \chi_\beta' &= \frac{-\delta(\Vec{\hat \mu}_1 \cdot \Vec{\hat{n}}) [\delta \gamma(\Vec{\hat \mu}_1 \cdot \Vec{\hat{\beta}} - \beta)]}{\delta\sqrt{1-\delta^2 \eta_0^2} \sqrt{1-(\Vec{\hat \mu}_1 \cdot \Vec{\hat{\beta}})^2}},\\
    \sin \chi_\beta' &= \frac{\mathrm{sgn}(r_{2,z}) (-\mu_{1,x}' \sin \phi + \mu_{1,y}'\cos \phi) }{\delta\sqrt{1-\delta^2\eta_0^2} \sqrt{1-(\Vec{\hat \mu}_1 \cdot \Vec{\hat{\beta}})^2}}.
\end{align}
Once the emerging polarization is obtained in the $\mathbf{B}(\Vec{\hat \mu}_2', \Vec{\hat n})$ basis, one can transform to $\mathbf{B}(\Vec{\hat \mu}_2, \Vec{\hat n})$ by doing the above rotation while substituting $\Vec{\hat \mu_1}$ with $\Vec{\hat \mu_2'}$,  $\Vec{\hat \mu_1'}$ with $\Vec{\hat \mu_2}$, and $\beta$ with $-\beta$.

Finally, we transformed from the local $\mathbf{B}(\Vec{\hat \mu}_2, \Vec{\hat n})$ to the global $\mathbf{B}(\Vec{\hat \mu}_2, \Vec{\hat \omega})$:
\begin{align}
    \cos \chi_2 &= \frac{n_\mathrm{z}- \eta \mu_{2,z}}{\sqrt{1 - \mu_{2,z}^2} \sqrt{1-\eta^2}}, \\
    \sin \chi_2 &= -\frac{\cos \zeta \sin(\psi_2 - \phi)}{\sqrt{1-\eta^2}},
\end{align}
where $\tan \psi_2 = \mu_{2,y}/\mu_{2,x}$.

\section{Approximating Chandrasekhar's H-functions} \label{sect:hfunc}

Chandrasekhar's H-functions can be approximated as 
\begin{equation} \label{eq:happrox}
    H(\mu) = \frac{1}{\prod_i\mu_\mathrm{i}} \frac{\prod_i(\mu+\mu_i)}{\prod_\alpha(1 + k_\alpha \mu)},
\end{equation}
where $\mu_i$ are the points of a quadrature rule of order $n$ in the interval $[-1, 1]$ and $k_\mathrm{\alpha}$ are the positive roots of the characteristic equation
\begin{equation}
    C(k) = 2\sum^n_{i=1} \frac{w_i \Psi(\mu_i)}{1-k^2\mu_i} - 1,
\end{equation}
where $w_i$ are the weights of the quadrature rule and $\Psi(\mu)$ is the characteristic polynomial associated with the H-function \citep{Chandrasekhar1960}.
The above equation has the same roots as the polynomial
\begin{align}
    P(k) &= C(k) F(k), \\
    F(k) &= \prod_{i=1}^n (1-k^2\mu_i).
\end{align}
The roots of the polynomial can be solved iteratively using the Aberth-Ehrlich method:
\begin{align}
    k_{\alpha,~j+1} &= k_{\alpha,~j} - \frac{N(k_{\alpha,~j})}{1-N(k_{\alpha,~j})\sum_{\beta\neq\alpha}\frac{1}{k_{\alpha,~j}-k_{\beta,~j}}},  \\
    N(k) &= \frac{P(k)}{P'(k)} = \frac{C(k)}{C(k)\sum\frac{2\mu_i^2k}{\mu_i^2k^2-1} + C'(k)},
\end{align}
where $k_{\alpha,~j}$ is the $j$th approximation of the $\alpha$th root.
The accuracy of the approximation can be assessed with the identity \citep{Chandrasekhar1947}
\begin{equation} \label{eq:h_accuracy}
    \int_0^1 H(\mu) \Psi(\mu) ~ \mathrm{d}\mu = 1 - \sqrt{1-2\int_0^1 \Psi(\mu)~\mathrm{d}\mu}.
\end{equation}

For the diffuse reflection from an electron-scattering atmosphere, the four relevant $H$-functions, $H_l$, $H_r$, $H_1$, and $H_2$, are defined by the characteristic polynomials
\begin{align}
    \Psi_l &=\frac{3}{4} \left( 1-\mu^2\right), &\Psi_r &= \frac{3}{8} \left( 1-\mu^2\right), \nonumber\\
    \Psi_1 &= \frac{3}{8}\left( 1-\mu^2\right) \left( 1+2\mu^2\right), &\Psi_2 &= \frac{3}{16} \left( 1+\mu^2\right)^2 \nonumber.
\end{align}
To approximate the $H$-functions, we calculated the roots of their characteristic equation for Gauss-Legendre quadrature points and weights of order 100.
The error of this approximation according to Eq. (\ref{eq:h_accuracy}) is $\lesssim10^{-9}$.
Evaluating the functions using Eq. (\ref{eq:happrox}) is computationally expensive, so we tabulated their values over a fixed grid and interpolated the results.

\end{document}